# Improved Phantom Cell deployment for Capacity Enhancement

Mahdi Ajamgard[1], Hamid Shahrokh Shahraki[2]

[1,2]Department of Electrical Engineering, University of Kashan, Kashan, Iran

[1] mehdiajamg@gmail.com

[2] shahrokh@kashanu.ac.ir

*Abstract*—Vastly increasing capacity and coverage demand in communication networks accompanied by energy efficiency challenge is getting attraction in research topics of this area. In this paper, an improved structure of Phantom cell heterogeneous networks (HetNets) is proposed to fulfil these requirements in the next generation cellular networks.It will be shown that the proposed orthogonal frequency-division multiple access (OFDMA) based structure can be employed in both indoor and outdoor environments by applying two individual frequency bands. Furthermore, the resource allocation problem of the proposed structure is investigated in downlink path. To this aim, a proper algorithm is presented in order to maximize the total throughput of all phantom cell users' equipments with regard to the protected minimum network capacity of the existing macrocell. To fulfil the objective goal, an iterative approach is employed in which OFDM subchannels and power transmitted by base stations are sequentially assigned and optimized at each step for every single frequency band. It is showed that the overall joint subchannel and power allocation algorithm converges to local maximum of the original designed problem. Performance improvement of the proposed algorithm is confirmed in both indoor and outdoor environments by numerical results.

## 1. INTRODUCTION

With the appearance of new popular wireless devices such as tablets, smart phones and other new wireless devices, there is a significant growth in wireless traffic demand [1]. Clearly the existing wireless cellular structures will not able to support the expected data rate and quality in the future, hence the necessity of reconsidering of the existing structures is undeniable.

Long Term Evolution-Advanced (LTE-A) is one of the best standards introduced by 3GPP to meet the huge traffic demand prospect [2]. Recent studies have been focused on various various features of LTE-A such as Massive MIMO, Carrier Aggregation,3-Dimention beamforming and Heterogeneous Networks to enhance the capacity [3-6].

Heterogeneous Networks are networks in which a large number of low power nodes distributed alongside a traditional Macrocell in order to increase the capacity and improve the network coverage. LTE-A introduced a special type of HetNets to meet the capacity leakage in indoor environments and hotspots known as femtocell HetNets[7-8]. In these cellular networks, the femtocells are small cells (low power nodes) that allowed frequency reuse with Macro cell equipments(MUE). By deploying femtocells, much higher spectral efficiency can be achieved within a specific geographical area. Moreover, since femto base stations are low power, capacity increasing reached in the green communication fashion.

However, femtocell user equipments (FUEs) cause interference in MUEs transmission, due to using the same frequency band used by MUEs. Extensive studies have been done on the context of optimum resource allocation in HetNets in order to maximize the total capacity with respect to the co-channel interference between FUEs and MUEs [9-12]. Nevertheless, it seems by operating one frequency band the transmission rates goal could not be reached.

On the other hand, there is a lack of capacity improvement solution for high demand outdoor zones. In outdoor Macro assisted small cells, users' motion in wide regions should be considered in addition to their connectivity.

In the current study we focus on using additional aspect of LTE-A, i.e. carrier aggregation, to the HetNets in order to increase the total throughput. According to carrier aggregation, various component carriers belong to different bands can be aggregated together so as to increase the transceivers' effective bandwidth and as a result improve the data rate [13].

A number of structures have been proposed to improve capacity of HetNets by carrier aggregation capability. The authors in [14], introduced a load-aware model for multi band carrier aggregated HetNets and explored different aspects of the proposed model such as biasing, base station density and transmit powers effects on the system's capacity.

Alternatively, DOCOMO proposed the phantom cells as a solution for improving the capacity of the outdoor HetNets' users by using carrier aggregation capability [15-16]. The proposed structure is based on using two different frequency in which the Macro cells operate in the lower frequency and phantom cells use higher frequency bands for their transmission. Moreover, in order to reduce frequent handover procedure, control plan and user plan are split for phantom cells. Clearly, as a result of using different frequency bands there isn't any interference between MUEs and phantom cell users (PUEs) in this configuration.

In this paper we introduce a new structure for Heterogeneous Networks. In fact this structure is an improved version of phantom cell HetNets which can be used in both indoor and outdoor environments. It is supposed that MUEs supported by the main carrier frequency and phantom cell users (PUEs) allowed frequency reuse with them by considering the tolerable co-channel interference constraint. Furthermore, PUEs can use additional frequency in order to get the maximum capacity.

Another contribution of this work is to investigate appropriate scheduling and resource allocation procedure for the proposed structure in downlink path. Accordingly, in the first step, the resource scheduling approach is formulated as an optimization problem for frequency F1. The objective function of the problem is to maximize the total throughput of all PUEs with regard to protected minimum rate, required for MUEs in each carrier, independently. By exploring the problem, a proper algorithm has been suggested for resource allocation in this step. In the second phase, another optimization problem has been formulated in order to maximize the PUEs' capacity on frequency band F2. Using the Lagrange dual decomposition method, a suitable algorithm has been extracted for simultaneous allocation of subcarriers and power for all phantom cell users.

The entire resource allocation procedure has been evaluated by simulation using realistic indoor/outdoor parameters. As simulation results verify, our suggested scenario enhanced the total capacity significantly compared with previous single frequency band HetNets. Moreover, it will be shown that a target rate can be reached expeditiously, with less power consumption which make it suitable from green communication point of view.

The remainder of this paper is organized as follows. In Section 2, the proposed scenario and its related formulation is discussed. In Section 3, a proper algorithm for resource allocation is presented for frequency band F1. Different stages of extracting a proper resource allocation algorithm for frequency band F2 are explored in section 4. section 5 analyzes complexity order of the proposed algorithms for practical applications. Performance evaluation of the proposed algorithm is verified by numerical experiments in Section 6 and finally, Section 7 concludes the paper.

## 2. SYSTEM MODEL AND PROBLEM FORMULATION

In this section we first present the suggested network structure and then formulate its parameters to derive a suitable resource allocation procedure. The network comprised of groups of phantom cells which are overlaid with a macrocell as illustrated in Fig.1. The macrocell is a conventional cell which supports its user with frequency band F1 based on OFDMA structure in downlink path. Macrocell base station is also used to handling of the phantom cells' control signaling via F1in order to minimize the number of PUEs handover.

Phantom cells are supplied with individual base stations which can be operate in two frequency bands (F1 and F2) with lower power and consequently small footprint. In frequency band F1 the phantom cells are allowed to use the same subcarriers with macrocell by considering the co-channel interference constraint. Frequency band F2 which is higher than F1 is used to enhance PUEs' throughput. Distinct streams of PUEs which are supplied by F1 and F2 can be synchronized through backhaul link among phantom and macro base stations besides carrier aggregation capability.

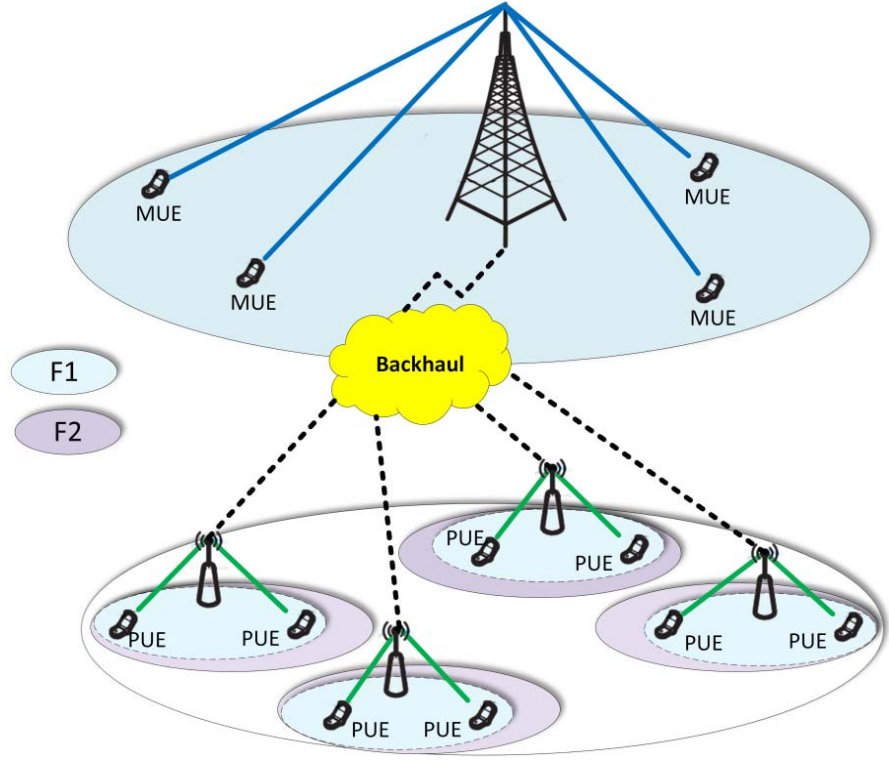

*Figure 1-Improved Phantom cell based HetNet, which MUEs supported by frequency band F1 and PUEs supplied with frequency bands F1 and F2.*

Let us denote the total number of subcarriers in the shared frequency band (F1) with $N_1$ and the number of subcarriers which used by phantom cells in frequency F2 with $N_2$. All cell base stations' are shown by set $A = \{0, 1, 2, \dots, M\}$ where index 0 is related to the macro cell and the others are phantom cells' index. Suppose that each cell has different number of users which are indicated by $K_m$ where $m \in A$. The capacity of $k$-th user on the $i$-th sub-carrier of $m$-th cell can be obtained by

$$r_{m,k}^i = ln\left(1 + \frac{h_{m,k}^i p_{m,k}^i}{I_{m,k}^i}\right) \quad (1)$$

Where $h_{m,k}^i$ is the channel gain between the $m$-th transmitter and $k$-th user's receiver on the $i$-th sub-carrier. $p_{m,k}^i$ denotes the $k$-th user's transmit power of $m$-th cell on the $i$-th sub-carrier and $I_{m,k}^i = \sum_{j \in A \setminus \{m\}} h_{j,k}^i p_{j,k}^i + N_0$ in which, $N_0$ is the additive white Gaussian noise variance.

As mentioned before in order to get the proper resource allocation, we should explore and formulate the problem for each carrier frequency individually. At the first step sub-carrier and power allocation should been done in such a way that the total capacity of phantom cells maximized while the macro cell users achieved their minimum required rates. To do so, we should solve the following optimization problem on frequency F1.

$$\text{maximize} \quad \sum_{m=1}^{M}\sum_{k=1}^{K_m}\sum_{i=1}^{N_1} \alpha_{m,k}^i \cdot r_{m,k}^i$$

$$\text{subject to} \quad \sum_{k=1}^{K_0}\sum_{i=1}^{N_1} \alpha_{0,k}^i \cdot r_{0,k}^i \geq R_{min} \quad\quad\quad\quad \textbf{P1}$$

$$\sum_{k=1}^{K_m} \alpha_{m,k}^i \leq 1 \quad \forall\, m,i \;\; and \;\; \alpha_{m,k}^i \in \{0,1\}$$

$$\sum_{i=1}^{N_1} \alpha_{m,k}^i \cdot p_{m,k}^i \leq P_m^{TH} \quad \forall\, m,k$$

$$0 \leq \alpha_{m,k}^i \cdot p_{m,k}^i \leq P_{m,k}^{i,MAX} \quad \forall\, m,i$$

In the problem **P1,** the second constraint is related to OFDMA nature of the supposed structure which indicates that each sub-carrier assigned to only one user in each cell. The next two constraints are related to the base stations' total power and spectral mask, respectively.

After finding a proper algorithm for resource allocation to meet constraints introduced in **P1**, at the second step we should solve the following optimization problem for frequency band F2. Note that after resource allocation based on **P1**, the macro cell users get their minimum required rates and since there is not any interference between F1 and F2, problem **P2** formulated to maximize only PUEs' rates. Therefore, in the following optimization problem in calculating $r_{m,k}^i$, $I_{m,k}^i$ which indicates co-channel interference among phantom cells will be $I_{m,k}^i = \sum_{j \in A \setminus \{m,0\}} h_{j,k}^i p_{j,k}^i + N_0$.

$$\text{maximize} \quad \sum_{m=1}^{M}\sum_{k=1}^{K_m}\sum_{i=1}^{N_2} \alpha_{m,k}^i \cdot r_{m,k}^i$$

$$\text{subject to} \quad \sum_{k=1}^{K_m} \alpha_{m,k}^i \leq 1 \quad \forall\, m,i \;\; and \;\; \alpha_{m,k}^i \{0,1\} \quad\quad \textbf{P2}$$

$$\sum_{i=1}^{N_1} \alpha_{m,k}^i \cdot p_{m,k}^i \leq P_m^{TH} \quad \forall\, k,i \; and \; m \in \{1,2,\dots,M\}$$

$$0 \leq \alpha_{m,k}^i \cdot p_{m,k}^i \leq P_{m,k}^{i,MAX} \quad \forall\, k,i \; and \; m \in \{1,2,\dots,M\}$$

The above problem is same as the problem **P1** except the first constraint which was related to the protected minimum capacity of the macrocell network. Nevertheless, this little difference lead to dissimilar procedure of solution for these problems. Therefore, we will explore **P1** and **P2** in individual sections.

### 3. EXTRACTING PROPER ALGORITHM FOR PROBLEM P1

The problem **P1** is a mixed integer nonlinear problem which is not feasible in general [17]. However, by some realistic assumptions and changes, an appropriate algorithm can be extracted to solve it. To this end, we first assume that there is not any phantom cell and determine feasibility of the problem by assigning optimal resource to macro cell.

Clearly, if the maximum achievable throughput by the macro cell is less than $R_{min}$, the problem is infeasible and it has not any solution.

Let us assume that the MUEs can achieve their minimum rate requirement and the problem is feasible. It has been shown that ignoring the phantom cells, maximum throughput of the macro cell can be achieved through the following procedure [14]:

- Optimal sub-carrier assignment for MUEs done as:

$$k_i^* = argmax\ h_{0,k}^i \quad k \in \{1,2,\dots,K_0\}\ and\ i \in N_1 \quad (2)$$

- By user assignment according to the above equation, the optimal power of each sub-carrier can be computed by the cap limited Water-filling algorithm as:

$$p_0^{(i)*} = \left[\frac{1}{\Psi} - \frac{N_0}{h_{0,k^*}^i}\right]_0^{P_0^{i,MAX}} \quad (3)$$

In which,

$$[x]_0^\beta = \begin{cases} \beta & if\quad x \geq \beta \\ x & if\quad 0 < x < \beta \\ 0 & if\quad x \leq 0 \end{cases} \quad (4)$$

And $\Psi$ is Lagrangian multiplier with the value:

$$\Psi = \sum_{i=1}^{N_1} p_0^{(i)*} = P_0^{TH} \quad (5)$$

After MUEs resource allocation according to the above method, if the computed total throughput $(R_0^*)$ is equal to the $R_{min}$ then there is not additional capacity for phantom cells and the problem's solution is finished. Otherwise, if $R_0^* > R_{min}$ then we should find an algorithm for phantom cell users in order to use the remaining capacity in the best manner.

Finding the optimum solution for simultaneously power allocation and user assignment for each phantom cell is very computed and impractical. Therefore, we separate these phases and find their jointly optimal point by an iterative manner. To do this, firstly, suppose that initial subcarrier assignment has been done for each phantom cell in any manner. Given a fixed subcarrier assignment, the following optimization problem should be solved to find the optimum power of $i$-th subcarrier in $m$-th phantom cell.

$$\max_{\{p_{m,k}^i\}} \sum_{m=1}^{M} \sum_{i=1}^{N_1} r_{m,k}^i$$

$$\text{subject to } \sum_{i=1}^{N_1} r_{0,k}^i \geq R_{min} \quad (6)$$

$$\sum_{i=1}^{N_1} p_{m,k}^i \leq P_m^{TH} \quad \forall \, m,k$$

$$0 \leq p_{m,k}^i \leq P_m^{i,MAX} \quad \forall \, m,i$$

Since, the first constraint of the above equation is not concave so the problem (6) is not convex. Therefore, in the present study, we use fmincon interior point function in MATLAB to find the optimal values of the above problem [18].

When the optimal solution of problem (6) is found, i.e. $\hat{p}_{m,k}^i$, subchannel assignment of PUEs can be updated as:

$$\alpha_{m,k}^{i\,*} = \begin{cases} 1 & \text{if } k = \arg\max_k r_{m,k}^i(\hat{p}_{m,k}^i) \\ 0 & \text{otherwise} \end{cases} \quad m \in \{1,2,\ldots,M\} \quad (7)$$

The above criteria denotes that, for any phantom cell subchannel $i$ assigned to the user which leads to the highest data rate on it.

The extracted procedures for updating power allocation and user assignment based on the above stages are repeated until convergence. Different steps of the proposed resource allocation procedures are summarized in algorithm 1.

**Algorithm 1** procedure for solving **P1**

---

- Compute Macrocell subcarriers' power assignment ($p_0^{(i)*}$), according to the equations (2) and (3) without regard to Phantom cells interference.

**IF** $\sum_{k=1}^{K_0} \sum_{i=1}^{N_1} \alpha_{0,k}^i \cdot r_{0,k}^i < R_{min}$

the problem has not any solution

**Else IF** $\sum_{k=1}^{K_0} \sum_{i=1}^{N_1} \alpha_{0,k}^i \cdot r_{0,k}^i = R_{min}$

The best solution is

$$p_0^{(i)*} = \left[\frac{1}{\Psi} - \frac{N_0}{h_{0,k^*}^i}\right]_0^{P_0^{i,MAX}} \quad for\ i \in N_1$$

$$P_m^i = 0 \quad for\ m \in \{1,2,\ldots,M\}$$

**Else**

- Initialize phantom cells' user assignment

**repeat** to satisfy problem conditions

- Allocate optimal power $\hat{p}_{m,k}^i$ for selected PUEs using fmincon function.

- Sort PUEs experiencing best channels andupdate subchannels assignment according to (7).

**until** convergence of $P_m^i$

---

# 4. EXTRACTING PROPER ALGORITHM FOR PROBLEM P2

The problem **P2** is again a mixed integer nonlinear problem which, finding its optimal solution need to exhaustive search among all possible states. Nevertheless, using a little change, the problem **P2** is converted to a convex problem and can be solved by the standard methods. We use the technique introduced in [19] and relax $\alpha_{m,k}^i$ to be a continuous real variable in the interval [0 1]. Now, by defining a new variable as $S_{m,k}^i = \alpha_{m,k}^i \cdot P_{m,k}^i$ the problem **P2** can be rewritten as below:

$$\text{maximize} \quad \sum_{m=1}^{M}\sum_{k=1}^{K_m}\sum_{i=1}^{N_2} \alpha_{m,k}^i \cdot \tilde{r}_{m,k}^i$$

$$\text{subject to} \quad \sum_{k=1}^{K_m} \alpha_{m,k}^i \leq 1 \quad \forall\, m,i$$

$$\alpha_{m,k}^i \in [0,1] \quad \forall\, m,i,k \tag{8}$$

$$\sum_{i=1}^{N_1} S_{m,k}^i \leq P_m^{TH} \quad \forall\, k \text{ and } m \in \{1,2,\ldots,M\}$$

$$0 \leq S_{m,k}^i \leq P_m^{i,MAX} \quad \forall\, k,i \text{ and } m \in \{1,2,\ldots,M\}$$

Where $\tilde{r}_{m,k}^i = \ln\left(1 + \frac{h_{m,k}^i S_{m,k}^i}{\alpha_{m,k}^i \cdot I_{m,k}^i}\right)$.

The objective function of the above problem is a concave function with respect to $S_{m,k}^i$ and $\alpha_{m,k}^i$ [17], therefore, as the constraints are all affine it has a unique optimal solution. We use Lagrange dual decomposition method to solve the problem by the following description.

The Lagrangian function of the problem (8) is:

$$L(\{\alpha_{m,k}^i\},\{p_m^i\},\boldsymbol{\lambda},\boldsymbol{\mu},\boldsymbol{\eta}) = \sum_{m=1}^{M}\sum_{k=1}^{K_m}\sum_{i=1}^{N_2} \alpha_{m,k}^i \tilde{r}_{m,k}^i - \sum_{m=1}^{M}\sum_{k=1}^{K_m} \lambda_{m,k}\left(\sum_{i=1}^{N_2} S_{m,k}^i - P_m^{TH}\right) \tag{9}$$

$$-\sum_{m=1}^{M}\sum_{k=1}^{K_m}\sum_{i=1}^{N_2} \mu_{m,k}^i(S_{m,k}^i - P_m^{i,MAX}) - \sum_{m=1}^{M}\sum_{i=1}^{N_2} \eta_m^i\left(\sum_{k=1}^{K_m} \alpha_{m,k}^i - 1\right)$$

Where, $\boldsymbol{\mu}$ and $\boldsymbol{\eta}$ are vectors of dual variables. Its Lagrange dual function can be written as:

$$g(\boldsymbol{\lambda},\boldsymbol{\mu},\boldsymbol{\eta}) = \max_{\{\alpha_{m,k}^i\},\{S_{m,k}^i\}} L(\{\alpha_{m,k}^i\},\{S_{m,k}^i\},\boldsymbol{\lambda},\boldsymbol{\mu},\boldsymbol{\eta}) \tag{10}$$

By appropriately changing the order in the summations of Eq.(9), it can be concluded that the optimization problem can be decomposed into $M \times N_2$ independent problems. Hence, the Lagrange dual function can be rewritten as:

$$L(\{\alpha_{m,k}^i\},\{S_{m,k}^i\},\lambda,\eta) = \sum_{m=1}^{M}\sum_{i=1}^{N_2} L_m^i(\{\alpha_{m,k}^i\},\{S_{m,k}^i\},\lambda,\mu,\eta) + \sum_{m=1}^{M}\sum_{k=1}^{K_m} \lambda_{m,k} P_m^{TH} \quad (11)$$

$$+ \sum_{m=1}^{M}\sum_{k=1}^{K_m}\sum_{i=1}^{N_2} \mu_{m,k}^i P_m^{i,MAX} + \sum_{m=1}^{M}\sum_{i=1}^{N_2} \eta_m^i$$

With the sub-problems optimization:

$$L_m^i(\{\alpha_{m,k}^i\},\{S_{m,k}^i\},\lambda,\eta) = \max_{\{\alpha_{m,k}^i\},\{S_{m,k}^i\}} \left\{ \sum_{k=1}^{K_m} \alpha_{m,k}^i \tilde{r}_{m,k}^i - \sum_{k=1}^{K_m} \lambda_{m,k} S_{m,k}^i - \sum_{k=1}^{K_m} \mu_{m,k}^i S_{m,k}^i - \sum_{k=1}^{K_m} \eta_m^i \alpha_{m,k}^i \right\} \quad \forall m,k$$

(12)

First, suppose that $\alpha_{m,k}^i$ be any given subcarrier assignment scheme for each phantom cell. Given a fixed $(\lambda,\eta,\mu)$, the objective function of the above problem is concave, therefore, its optimal solution can be extracted using Karush-Kuhn-Tucker (KKT) conditions as:

The derivation of Eq.(12) regarding $S_{m,k}^i$ is:

$$\frac{\partial L_m^i(\{\alpha_{m,k}^i\},\{S_{m,k}^i\},\lambda,\eta)}{\partial S_{m,k}^i} = \frac{\alpha_{m,k}^i h_{m,k}^i}{\alpha_{m,k}^i I_{m,k}^i + h_{m,k}^i \hat{S}_{m,k}^i} - \mu_{m,k}^i - \lambda_{m,k} \quad (13)$$

By setting the above derivative equal to zero and considering $\mu_{m,k}^i \geq 0$ and $S_{m,k}^i = \alpha_{m,k}^i \cdot P_{m,k}^i$, the optimal power for the user $k$ in phantom cell $m$ at subchannel $i$ can be obtained as:

$$\hat{p}_{m,k}^i = \frac{\hat{S}_{m,k}^i}{\alpha_{m,k}^i} = \left[\frac{1}{\lambda_{m,k}} - \frac{I_{m,k}^i}{h_{m,k}^i}\right]_0^{P_m^{i,MAX}} \quad \forall m,i \quad (14)$$

Now, considering the fact that each subcarrier can be assigned to utmost one user in each phantom cell, optimal subcarrier policy can be derived by substituting Eq.(14) in Eq.(12) and searching among all users. As a result:

$$\hat{\alpha}_{m,k}^{i\,*} = \max_{k}\left\{\ln\left(1 + \frac{h_{m,k}^i \hat{S}_{m,k}^i}{I_{m,k}^i}\right) - \lambda_{m,k}\hat{S}_{m,k}^i - \mu_{m,k}^i \hat{S}_{m,k}^i - \eta_m^i\right\} \quad (15)$$

And since $\eta_m^i \geq 0$, it can be conclude from Eq.(15) that, Sub-channel $i$ is assigned to user $k$ with the largest $H_{m,k}^i$ in phantom cell $m$ that:

$$H_{m,k}^i = \ln(1 + \frac{h_{m,k}^i \hat{p}_{m,k}^i}{I_{m,k}^i}) - \lambda_{m,k}\hat{p}_{m,k}^i - \mu_{m,k}^i \hat{p}_{m,k}^i \qquad (16)$$

The remaining task is to discover an optimal value for $\lambda_{m,k}$ and $\mu_{m,k}^i$ as the equations (14) and (16) don't depend on the dual variable $\eta_m^i$. Toward this end, the dual problem of Eq.(10) should be addressed and the following problem is to be solved:

$$\min_{\lambda,\mu} g(\boldsymbol{\lambda}, \boldsymbol{\mu}, \boldsymbol{\eta}) \\ \text{subject to } \boldsymbol{\lambda}, \boldsymbol{\mu} \geq 0 \qquad (17)$$

As the Lagrange dual problem is always convex, the above problem can be solved by any iterative method. In the present study, subgradient method is used to iteratively update the values of $\lambda_{m,k}$ and $\mu_{m,k}^i$ until convergence. The updates can be performed as:

$$\lambda_{m,k}^{(t+1)} = \left[\lambda_{m,k}^{(t)} - \beta_1^{(t)}\left(P_m^{TH} - \sum_{i=1}^{N_2} s_{m,k}^i\right)\right]^+, \forall m,k \qquad (18)$$

$$\mu_{m,k}^{i,(t+1)} = \left[\mu_{m,k}^{i,(t)} - \beta_2^{(t)}\left(P_m^{i,MAX} - S_{m,k}^i\right)\right]^+, \forall m,k,i \qquad (19)$$

Where $\beta_1^{(t)}$ and $\beta_2^{(t)}$ are the step size and chosen to be sufficiently small. Step sizes can be constant value or chosen in diminishing rules with respect to iteration order "$t$" [20].

    The main steps of the extracted procedure of resource allocation on frequency F2 are outlined in algorithm 2. Note that, by using algorithms 1 and 2 even there was not any resource for the phantom cell users in frequency F1, for sure they can gain some throughput related to the F2 usage.

**Algorithm 2** procedure for solving **P2**

---

**Initialize** $\lambda, \mu$ and **set** t= 1

**Initialize** all phantom cells' user assignment $\alpha_{m,k}^i$

**Repeat**

- Find $\hat{p}_{m,k}^i$ according to (14)
- Update $\hat{\alpha}_{m,k}^{i\,*}$ according to (15) and (16)
- Update $\lambda, \mu$ according to (18) and (19)

**Until** convergence of $\lambda, \mu$

---

## 5. COMPEXITY ANALYSIS

In this section, the computational complexity of the proposed algorithms is analyzed. Let us assume that the mean number of users in each phantom cell is equal to $K$, the number of phantom cells $M$ and the number of available subcarriers on frequency bands F1 and F2 are $N_1$ and $N_2$, respectively. In the algorithm1, finding the optimal solution using equations (2) and (3) has the complexity of $\mathcal{O}(N_1 log_2 N_1)$. For a fixed user assignment finding the optimal power allocation based on using the fmincon function has the extreme complexity of $\mathcal{O}(N_1^3 M^3)$. Therefore, if we suppose that the iterative structure of algorithm1 converges after $\Delta_1$ iterations, then the overall complexity of the algorithm1 is $\mathcal{O}(KN_1^3 M^3 \Delta_1)$.

In the algorithm2, with a fixed, $\mu$ and $\eta$, equation (14) needs $\mathcal{O}(N_2 log_2 N_2)$ operations to find its optimal soloution. Consequently, finding the best user assignment and power allocation has the computational complexity of $\mathcal{O}(KN_2 log_2 N_2)$. On the other hand, if the subgradient method used to update $\lambda$ and $\mu$ needs $\Delta_2$ iterations for convergence, the total complexity order of algorithm2 will be $\mathcal{O}(\Delta KN_2 log_2 N_2)$. It has been shown that $\Delta_2$ is polynomial function of $MN_2$ [21].

As it can be seen, both of the proposed algorithms have polynomial complexity order of system's parameters, which make them suitable for practical applications. Besides, it is important to note that, since these algorithms are used to determine the scheduling procedure in two different frequencies, therefore, they can be computed simultaneously.

# 6. PERFORMANCE EVALUATION

Evaluation of the proposed resource allocation algorithm in terms of simulation results is given in this section. The simulations are carried out for both indoor and outdoor application of the proposed structure as shown in Fig.2. The coverage radius of macrocell and phantom cell is 1000m and 50m, respectively, where phantom cell BSs is serving as an indoor hotspot. In the other scenario, where phantom cell BSs serve outdoor users, macro cell has a coverage radius of 1000m and phantom cells have a radius of 250m.

In the both scenarios, there are 4 phantom cells placed uniformly inside an existing macro cell. MUEs ($k_0 = 10$) and PUEs ($k_i = 5, i = \{1,...,4\}$) are placed randomly with uniform distribution in angle and Gaussian distribution in distance inside their corresponding cells. MUEs and PUEs are sharing spectrum within frequency band F1 whereas PUEs benefit from another frequency band, F2, as well. Each frequency band has N=8 sub-channels with bandwidth of BW=180 KHz. Maximum transmit power of macrocell BS is set 47dBm and of phantom cells is 23dBm for indoor scenario and 30dBm for outdoor scenario. OFDM power mask is $P^{MAX} = P^{TH}/N$ and noise power is equal to $\sigma^2 = BW.N_0$ where $N_0 = -174dBm/Hz$ is the AWGN power spectral density.

The wireless model is composed of path loss (PL), shadowing (S) and frequency-flat Rayleigh fading (F). Different types of path losses of different links are presented in Table 1. $L_{wall}$ is wall penetration and equals to 10dB. The Rayleigh fading gains are modeled as unit-mean exponentially distributed random variables. Shadowing fluctuations have zero-mean log-normal distribution with variance of 10dB.

Table 2 summarizes simulation parameters which are used for both scenarios in our evaluation.

*Table 1- Calculation of different link path losses*

| Link Type | Path loss |
|---|---|
| MBS to MUE | $PL = 15.3 + 37.6 \log_{10} d$ |
| MBS interfering PUE | $PL = 15.3 + 37.6 \log_{10} d + L_{wall}$ |
| PBS to indoor PUE | $PL = 38.46 + 20 \log_{10} d$ |
| PBS to outdoor PUE | $PL = 15.3 + 37.6 \log_{10} d$ |
| PBS interfering MUE | $PL = 38.46 + 20 \log_{10} d + L_{wall}$ |

*Table 2- Simulation setup summary*

| Parameter | Macro cell | Phantom cell |
|---|---|---|
| Radius | 250m / 1000m | 50m / 250m |
| UE | 10 MUEs | 4 * 5 PUEs |
| Sub-channels | $N_1 = 8$ | $N_1 = N_2 = 8$ |
| SC-Bandwidth | 180KHz | 180KHz |
| Power | 47dBm@F1 | 23dBm@F1, F2 / 30dBm@F1, F2 |
| Spectral Mask | $P^{MAX} = P^{TH}/N$ | |
| Channel Model | $h_{m,k}^i = S_{m,k}^i . F_{m,k}^i . PL_{m,k}^{-1}$ | |
| Monte-Carlo Averages | 80 | |

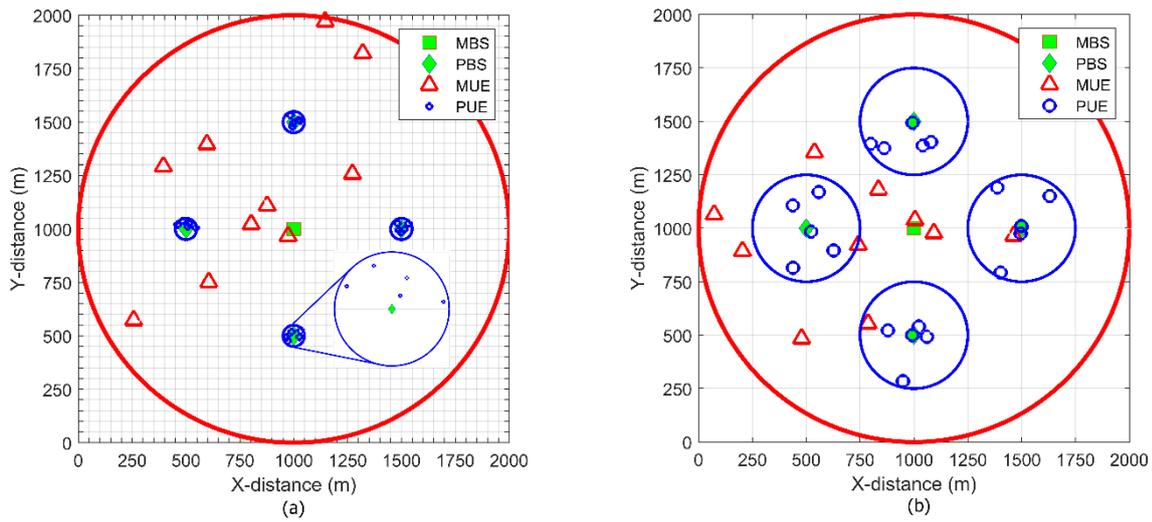

*Figure2- Numerical example of network deployment geometry: (a) Indoor scenario, (b) Outdoor scenario*

Fig. 3 shows sub channel assignment for indoor and outdoor scenarios according to Eq (7). Notably, some users occupied most of sub channels in their corresponding cell (*e.g.* PUE5 of cell 1 and PUE1 of cell 3 in Fig3.b). One possible explanation is highly favorable conditions of that channels for these UEs in our randomly generated distribution. Also there are channels that have not been assigned to any UEs; UEs experience poor conditions in these channels.

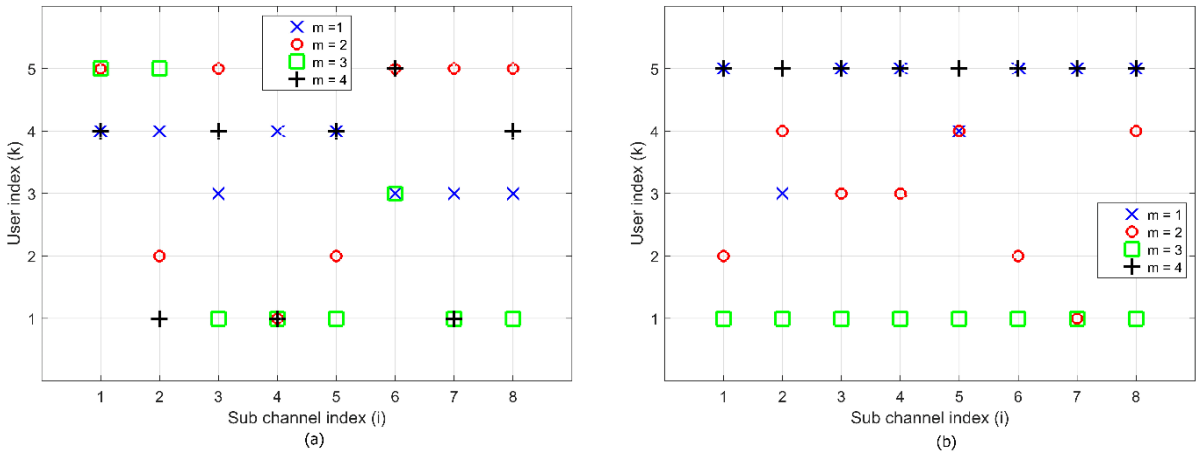

*Figure 3- Subchannel assignment for PUEs: (a) indoor, (b) outdoor scenario. Users belonging to same cells represented by same marker type.*

Fig. 4 shows that the average power allocated per user is saturated with maximum accessible power for each user. It can be seen in this figure, the convergence speed of power allocation scheme based on the subgradient method is expeditious and allocated power for indoor and outdoor scenarios settle to stable solutions after a few iterations.

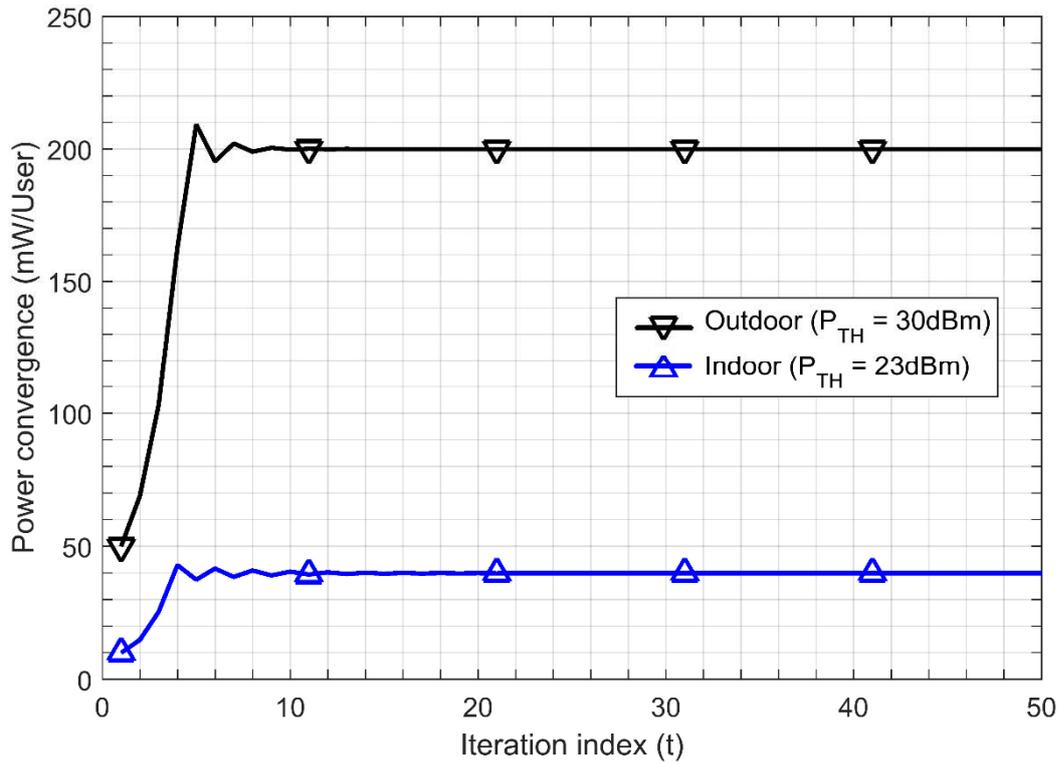

*Figure 4- The convergence in terms of average power consumed per phantom user over the number of iterations for indoor and outdoor scenarios*

Fig. 5 shows comparison of capacity obtained through the fmincon algorithm and subgradient method with same simulation conditions. In spite the fact that the fmincon algorithm has superior performance than subgradient method, it can be traded off with lower computational complexity of subgradient method. For a dual core CPU computer supported with 4GB of RAM, it takes around 40 seconds to solve an iteration of the fmincon method compared to 1.8 seconds time consumed for solving one iteration of the subgradient method. Outdoor performance is inferior than capacity obtained by indoor scenario, but it is still acceptable.

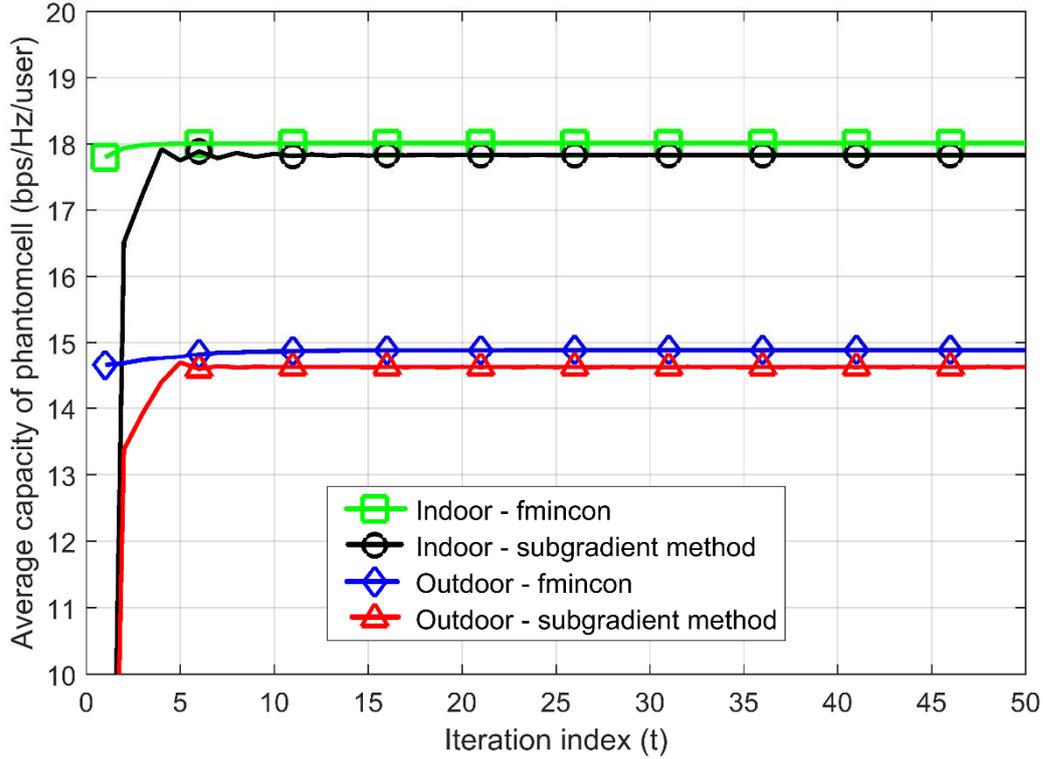

*Figure 5- The convergence comparison of fmincon and subgradient method in terms of average capacity of phantom users over the number of iterations for indoor and outdoor scenarios*

Finally, Fig. 6 shows the average phantom user capacity over minimum capacity demanded by macro cell for indoor and outdoor scenarios. Capacity of F1 decreases by increase of macro cell minimum required capacity. Macro cell higher capacity means more power consumption by macro cell base station. As a consequence, total interference of macro cell base station on phantom users sharing F1 frequency band became higher and realizable capacity for phantom users decreased. In frequency band F1, for values higher than a certain point of required capacity by macro cell, there will be no capacity remained for phantom users to claim. Nevertheless, phantom users do not suffer from interference caused by macro cell in frequency band F2. Therefore, phantom users' QoS is guaranteed by carrier aggregation.

Maximum power for Indoor phantom BS is 23dBm and coverage area radius is 50m; For outdoor scenario these parameters set 30dBm and 250m, respectively. Fig. 6 shows that outdoor capacity is lower than indoor capacity. One reason is range expansion factor of coverage area is higher than power increment factor from indoor scenario to outdoor scenario; Therefore, obtained capacity in outdoor phantom cell is lower.

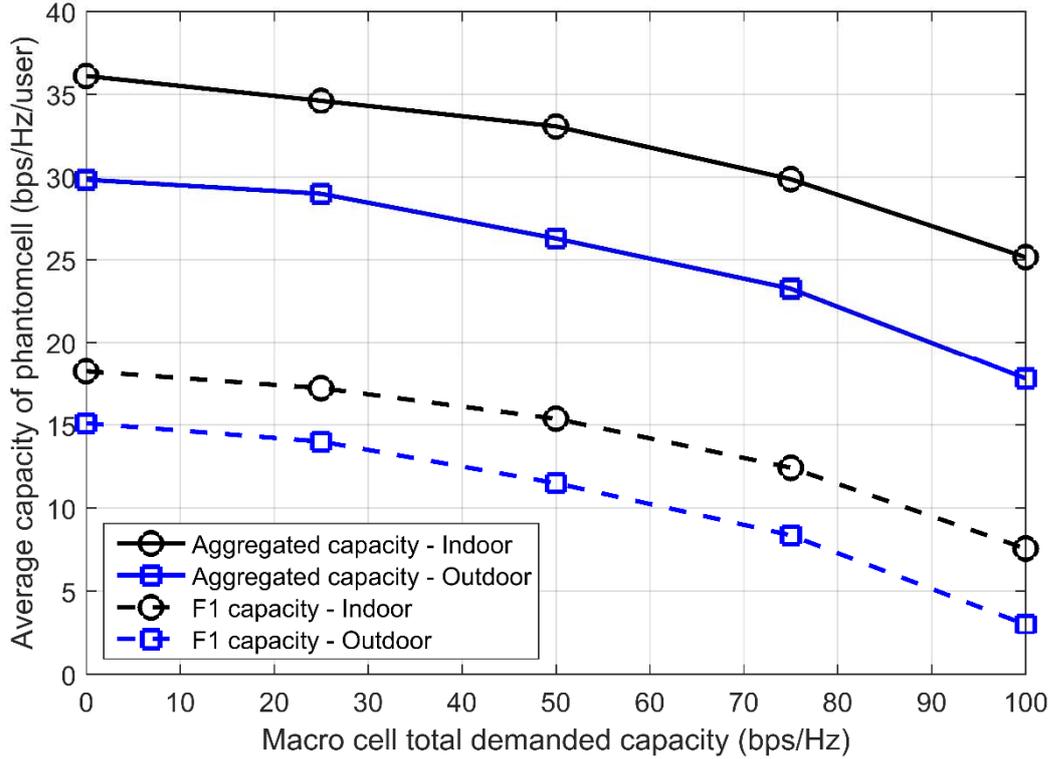

*Figure 6- Average capacity of phantom users over minimum capacity guaranteed for macrocell*

## 7. CONCLUSIONS

In this paper we introduced an improved structure of phantom cell based HetNets for indoor and outdoor applications. We used carrier aggregation capability of LTE-A standard and by applying two different frequency band tried to enhance PUEs total throughput. In this context, the resource allocation problem of the proposed structure has been explored in downlink path. The problem has been formulated as two individual optimization problems for different frequency bands. For frequency band F1, the optimization objective is to maximize phantomcells' throughput while overlooking co-channel interference to protect macrocell's minimum required throughput. Including an in-depth theoretical analysis, a proper algorithm has been suggested to fulfill the goal. Dual Lagrange method has been used to extract efficient algorithm for resource allocation problem on frequency band F2. The effectiveness of the proposed algorithms which are shown by numerical experiments, makes our research a competitive algorithm for implementing in a practical cellular communication scenario.